\documentclass{article}
\usepackage{graphicx} 
\usepackage{amsmath}
\usepackage{float}
\usepackage{tabularx}
\usepackage{amssymb}
\usepackage{caption}
\usepackage{subcaption}
\usepackage{hyperref}
\usepackage{adjustbox}
\usepackage{authblk}
\usepackage{float}
\usepackage{booktabs}
\usepackage{xcolor}
\title{A Dual-Path Framework with Hotspot-Guided Fusion for Three-Dimensional CT-to-PET Synthesis in Head and Neck Cancer}
\author[1]{Mohd Maaz Khan}
\author[1,2]{Oluwaseyi Oderinde}
\affil[1]{\small Advanced Molecular Imaging in Radiotherapy (AdMIRe) Research Laboratory, School of Health Sciences, Purdue University, West Lafayette, IN, 47907, USA}
\affil[2]{\small Department of Radiation Oncology, Indiana University School of Medicine, Indianapolis, IN 46202, USA}

\date{May 2026}

\begin{document}

\maketitle

\begin{abstract}

\textbf{Objective:} \textsuperscript{18}F-FDG PET/CT plays a central role in staging, treatment planning, and response assessment for head and neck cancer by providing functional information that complements anatomical CT imaging. However, PET acquisition requires radiotracer administration, specialized infrastructure, and additional cost, limiting its availability for repeated imaging. We present a proof-of-concept deep learning framework for synthesizing PET-like images directly from routine CT scans with the goal of providing complementary metabolic information that may support imaging triage and clinical decision support rather than replace diagnostic PET.

\textbf{Approach:} Forty-four patients from the publicly available QIN-\newline HEADNECK dataset were retrospectively analyzed using five-fold cross-validation. We propose a fully three-dimensional dual-path architecture consisting of (i) a regression U-Net optimized for voxel-wise quantitative SUV estimation and (ii) a conditional generative adversarial network optimized for realistic PET texture. Their outputs are integrated using hotspot-guided Laplacian pyramid blending, allowing quantitative information from the regression pathway to be preserved within metabolically active regions while leveraging adversarial texture synthesis elsewhere.

\textbf{Main Results:} The proposed framework achieved a mean absolute error of 0.00395, PSNR of 39.19 dB, and SSIM of 0.9634 on reconstructed three-dimensional PET volumes. Qualitative evaluation demonstrated accurate localization of many FDG-avid lesions while producing anatomically realistic background texture. Consistent with previous CT-to-PET synthesis studies, the principal limitation was systematic underestimation of SUV within highly metabolically active tumor regions.

\textbf{Conclusion:} A dual-path three-dimensional CT-to-PET synthesis framework can generate PET-like volumes that combine quantitative stability in regions of elevated uptake with realistic background appearance. Although the current model is not sufficiently accurate for quantitative PET interpretation or clinical decision-making, the results demonstrate the feasibility of estimating metabolically informative imaging from routine CT alone. Future work will focus on improving quantitative accuracy in high-uptake lesions and validating the framework on larger multi-institutional cohorts to determine its potential role in clinical workflow support.
\end{abstract}

\textbf{Keywords:}  CT-to-PET synthesis, FDG-PET, head and neck cancer, generative adversarial network, U-Net, Laplacian pyramid blending, virtual molecular imaging

\section{Introduction}

\textsuperscript{18}F-fluorodeoxyglucose positron emission tomography/computed tomography (\textsuperscript{18}F-FDG PET/CT) is an essential imaging modality for the management of head and neck squamous cell carcinoma (HNSCC). By combining anatomical information from CT with measurements of glucose metabolism, PET/CT improves tumor staging, radiotherapy target delineation, treatment response assessment, and recurrence detection.\cite{dong2019synthetic}. Unlike CT, which primarily depicts tissue morphology, FDG-PET characterizes tumor biology by identifying regions of increased metabolic activity, providing complementary information that is often unavailable from anatomical imaging alone \cite{almutairi2025application, almutairi2026metabolic}.

Despite its clinical importance, PET imaging remains substantially more expensive and less accessible than CT because it requires radiotracer administration, specialized imaging systems, and dedicated radiopharmaceutical infrastructure. Consequently, CT is routinely acquired for nearly every patient undergoing radiotherapy, whereas PET is often unavailable because of cost, logistics, or limited access. These challenges have motivated increasing interest in computational methods that estimate PET-like metabolic information directly from routinely acquired CT images.\cite{chandrashekar2022virtual} \cite{bi2017synthesis}.

Recent advances in deep learning have made cross-modality image synthesis increasingly feasible. Several studies have demonstrated that convolutional neural networks \cite{bencohen2017virtual, chandrashekar2022virtual}, conditional generative adversarial networks (cGANs) \cite{bencohen2019cross, bi2017synthesis, chai2025synthesizing, li2023fully, xu2025whole}, and more recently diffusion models \cite{nguyen2025ct, mahdi2026ct, xie2024synthesizing} can generate synthetic PET images from CT with encouraging visual quality. However, predicting metabolism from anatomy remains fundamentally challenging because FDG uptake reflects complex biological processes—including glucose metabolism, proliferation, inflammation, and hypoxia that are only partially represented by CT. This challenge is particularly pronounced in head and neck cancer because of complex anatomy, heterogeneous tumor appearance, and physiological FDG uptake in adjacent normal tissues\cite{boellaard2015fdg}.

\begin{figure}[H]
    \centering
    \includegraphics[width=0.85\textwidth]{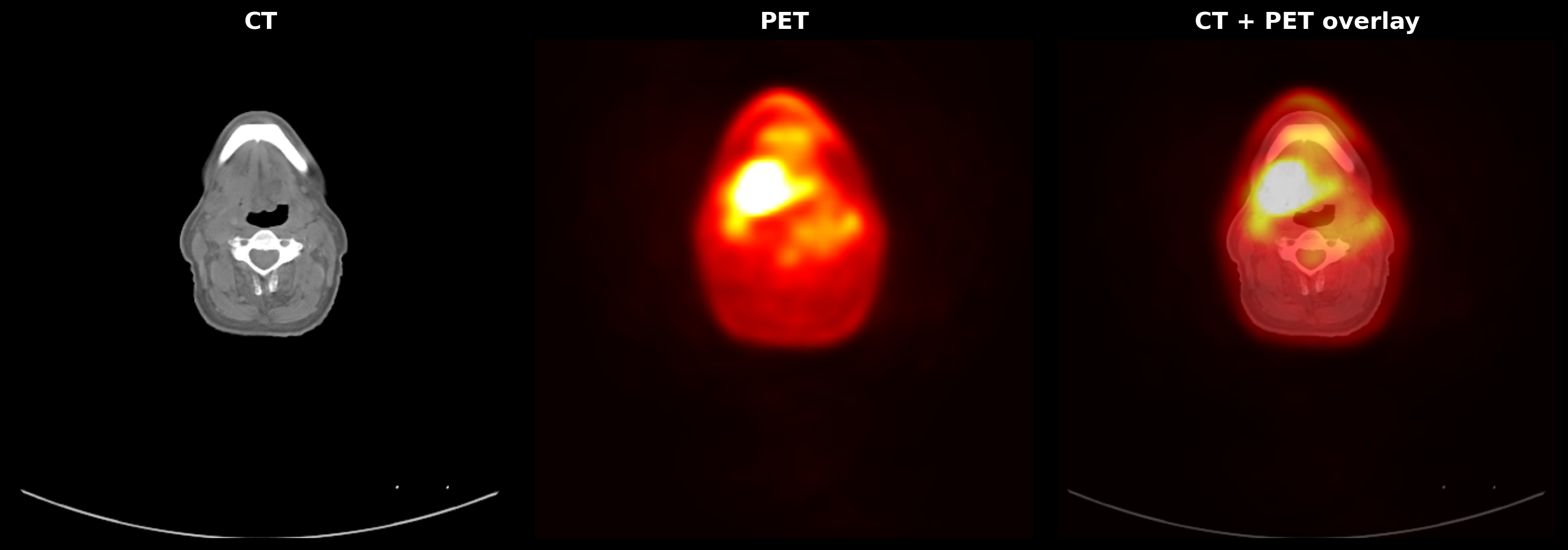}
    \caption{Representative axial slice from a head-and-neck FDG-PET/CT study. Left: CT showing anatomical detail. Middle: FDG-PET on a hot colormap, where bright/yellow voxels correspond to regions of high $^{18}$F-FDG uptake (high SUV) such as the metabolically active tumor. Right: PET overlaid on CT.}
    \label{fig:pet_motivation}
\end{figure}

Current CT-to-PET synthesis methods also face an inherent trade-off between quantitative accuracy and perceptual realism. Regression-based models generally provide more reliable voxel-wise standardized uptake value (SUV) estimation but often generate overly smooth images. In contrast, adversarial models produce sharper and more realistic PET textures but may distort quantitative SUV measurements or generate spurious uptake patterns \cite{zhu2017unpaired}. Consequently, many existing methods systematically underestimate SUV within highly metabolically active lesions, limiting their clinical applicability despite producing visually convincing images.

We hypothesize that quantitative SUV fidelity and perceptual realism should be optimized independently and subsequently integrated rather than being learned simultaneously by a single network. Based on this hypothesis, we developed a fully three-dimensional dual-path CT-to-PET synthesis framework that combines a regression U-Net for quantitative SUV prediction with a conditional GAN for realistic image synthesis. Their outputs are fused using a hotspot-guided three-dimensional Laplacian pyramid blending strategy that preserves quantitative information within predicted high-uptake regions while leveraging adversarial texture synthesis elsewhere.

The principal contributions of this work are fourfold. First, we propose a fully volumetric dual-path framework for CT-to-PET synthesis in head and neck cancer that jointly addresses quantitative SUV estimation and perceptual image quality. Second, we introduce a hotspot-guided three-dimensional Laplacian pyramid blending strategy that selectively combines regression- and adversarial-based predictions according to their complementary strengths. Third, we evaluate the proposed framework on the publicly available QIN-HEADNECK dataset using patient-level five-fold cross-validation and quantitative image quality assessment. Finally, we position synthetic PET as a proof-of-concept approach for estimating metabolically informative imaging from routine CT, providing a foundation for future investigation of CT-derived functional imaging as an adjunct to, rather than a replacement for, diagnostic PET/CT.

\section{Methods}

\subsection{Dataset}

The proposed framework was developed and evaluated using the publicly available QIN-HEADNECK dataset available through The Cancer Imaging Archive (TCIA)\cite{chandrashekar2022virtual}. The dataset comprises co-registered planning CT and pre-treatment 18F-FDG PET/CT images acquired from patients with head and neck squamous cell carcinoma (HNSCC), together with the associated DICOM metadata required for quantitative PET analysis. Additionally, the cohort included patients with locally advanced HNSCC representing multiple primary tumor sites and disease stages, reflecting the clinical heterogeneity encountered in routine radiotherapy planning. 

One patient (QIN-HEADNECK-01-0213) was excluded because the DICOM metadata required for standardized uptake value (SUV) conversion were incomplete. The final study cohort therefore consisted of 44 patients with paired CT and PET imaging available for model development and evaluation.

To provide reference regions of metabolically active disease for qualitative evaluation, gross tumor volumes (GTVs) were delineated on the PET images using a 45\% maximum-SUV contour threshold, a commonly adopted semi-automatic segmentation approach for FDG-avid head and neck tumors. These contours were used only for visualization and qualitative assessment of lesion localization and were not provided as inputs to the proposed synthesis framework during either training or inference.

The dataset was divided using patient-level five-fold cross-validation, ensuring that images from the same patient were never present in both the training and validation sets. All quantitative results reported in this study were obtained by averaging performance across the cross-validation folds unless otherwise specified.

\subsection{Imaging Preprocessing}
\textbf{CT Preprocessing.}
The planning CT images were preprocessed to standardize image intensity and reduce variability across patients before network training. CT intensities were clipped to the range $[-160, 240]$, capturing soft tissue structures relevant to head and neck while suppressing image regions unlikely to contribute meaningful metabolic information, such as air cavity. The lower bound excludes air cavities and most fat, which have very low FDG uptake. The upper bound excludes dense cortical bone ($>400$ HU). The windowed intensities were subsequently linearly normalized to the range $[0,1]$, providing a consistent input scale for network optimization.
\begin{equation}
    \tilde{x}_{CT} = \frac{\mathrm{clip}(x_{CT},\,-160,\,240)+160}{400}
    \label{eq:ct_norm}
\end{equation}
where $x_{\text{CT}}$ denotes the original CT intensity in Hounsfield units.

\textbf{PET Preprocessing.}
PET images were converted from activity concentration (Bq/mL) to standardized uptake value (SUV) using patient-specific injected dose and body-weight information extracted from the DICOM metadata. SUV normalization provides a quantitative measure of tracer uptake that is comparable across patients and is the standard representation used in clinical PET analysis.
\begin{equation}
    \text{SUV}(x) = \frac{C_{\text{voxel}}}{D_{\text{corrected}} / W}
    \label{eq:suv}
\end{equation}
where $C_{\text{voxel}}$ is the voxel activity concentration (Bq/mL), $D_{\text{corrected}}$ is the decay-corrected injected dose (Bq), and $W$ is patient body weight (g). To reduce the influence of extreme outliers while preserving clinically relevant uptake values, SUV values were clipped to the range $[0,20]$ and normalized to $[0,1]$, giving our network a consistent, patient-independent scale.

\textbf{Volumetric Patch Generation.}
The original CT and PET volumes varied substantially in axial extent and exceeded the memory limitations of end-to-end three-dimensional network training. Therefore, each volume was divided into overlapping three-dimensional patches consisting of 32 axial slices with an in-plane resolution of 256 × 256 pixels.

During inference, synthetic PET patches were reconstructed into full patient volumes using a sliding-window strategy with 50\% overlap. Predictions from overlapping regions were combined using cosine-weighted averaging, reducing boundary artifacts and ensuring smooth transitions between adjacent patches while preserving volumetric continuity.

All preprocessing operations were performed identically for both the training and validation datasets to ensure consistency throughout model development.

\subsection{Overall Framework}
The proposed framework was designed to address a fundamental limitation of existing CT-to-PET synthesis methods: the competing objectives of quantitative accuracy and perceptual image realism. Regression-based models generally provide stable voxel-wise SUV estimation but tend to generate overly smooth PET images. Conversely, adversarial models produce realistic image texture but may distort SUV values or introduce spurious uptake patterns, particularly within metabolically active tumor regions. Optimizing a single network to satisfy both objectives simultaneously is therefore challenging.

\begin{figure}[H]
    \centering
    \includegraphics[width=\textwidth]{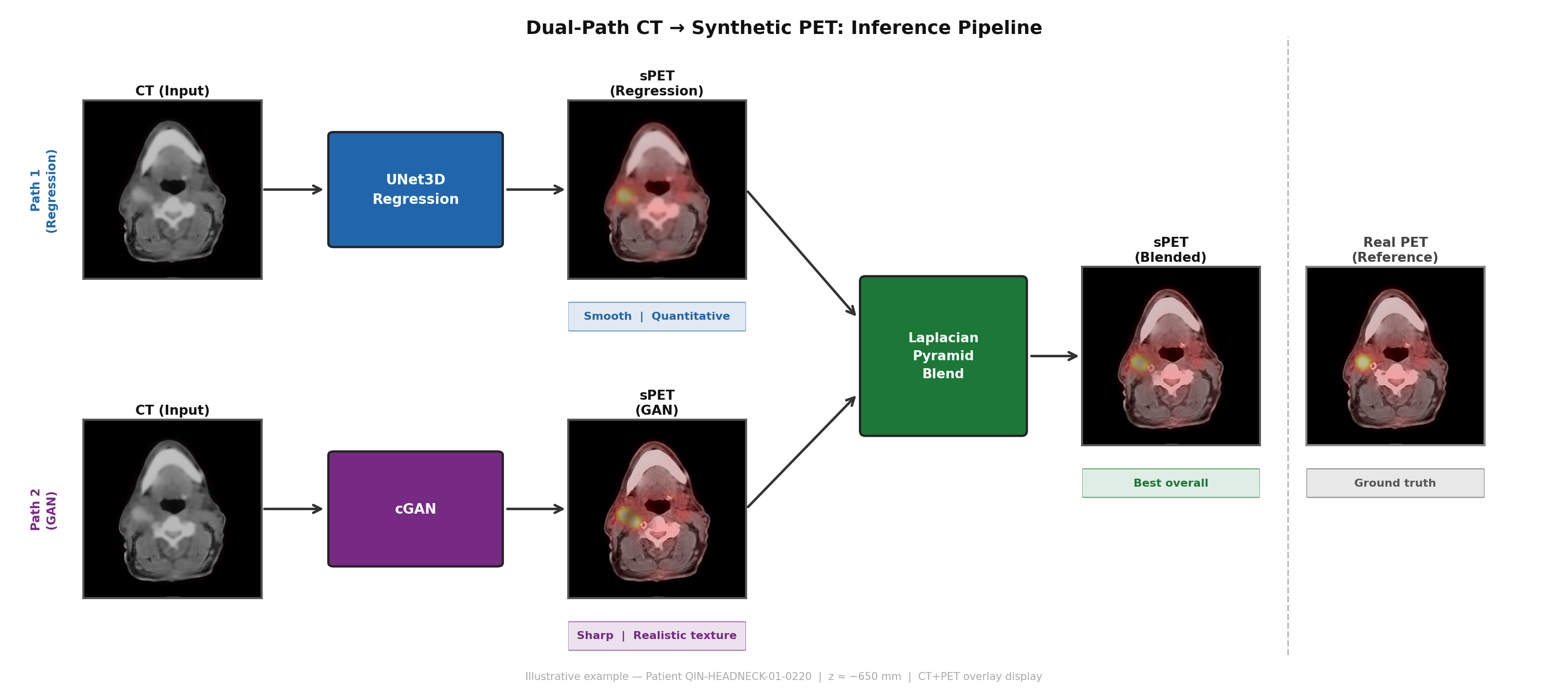}
    \caption{Dual-path CT-to-FDG-PET synthesis pipeline}
    \label{fig:pipeline}
\end{figure}

To address this limitation, we propose a dual-path three-dimensional CT-to-PET synthesis framework that independently optimizes quantitative prediction and perceptual realism before selectively combining their complementary strengths. An overview of the proposed framework is illustrated in Figure~\ref{fig:pipeline}. Whereas the regression pathway minimizes quantitative reconstruction error, the adversarial pathway learns the distribution of realistic PET appearance. These objectives are complementary rather than redundant, forming the basis of the proposed dual-path architecture.

Given an input planning CT volume, two parallel synthesis pathways are executed simultaneously. The first pathway is a regression-based three-dimensional U-Net that learns a direct voxel-wise mapping from CT to PET and is optimized to accurately estimate SUV throughout the imaging volume. The second pathway is a three-dimensional conditional generative adversarial network (cGAN) that generates PET images with realistic texture and local structural characteristics by learning the underlying PET image distribution.

Rather than averaging the outputs of both networks, the proposed framework performs region-specific fusion using a hotspot-guided three-dimensional Laplacian pyramid blending strategy. A soft metabolic mask is first generated from the regression prediction to identify regions likely to exhibit elevated FDG uptake. Within these metabolically active regions, the framework preferentially preserves the regression prediction because quantitative SUV accuracy is of primary clinical importance. Outside the hotspot region, the adversarial prediction is emphasized to preserve realistic PET texture and anatomical appearance. Frequency-selective Laplacian blending ensures smooth transitions between the two reconstructions while minimizing boundary artifacts \cite{bencohen2019cross}.

The final output is a fully reconstructed three-dimensional synthetic PET volume that combines the quantitative stability of the regression pathway with the perceptual realism of the adversarial pathway. The individual components of the framework, including the regression network, adversarial network, and hotspot-guided fusion module are described in the following sections.

\subsection{Regression Path}
The objective of the regression pathway is to generate an accurate voxel-wise estimate of the SUV from the input CT volume. Within the proposed dual-path framework, this branch serves as the quantitative component of the synthesis pipeline by learning the direct mapping between anatomical CT features and PET tracer uptake. Unlike adversarial learning, which primarily encourages perceptually realistic image generation, the regression pathway is optimized to preserve quantitative fidelity, particularly within metabolically active tumor regions where accurate SUV estimation is clinically important.

\begin{figure}[H]
    \centering
    \includegraphics[width=\textwidth]{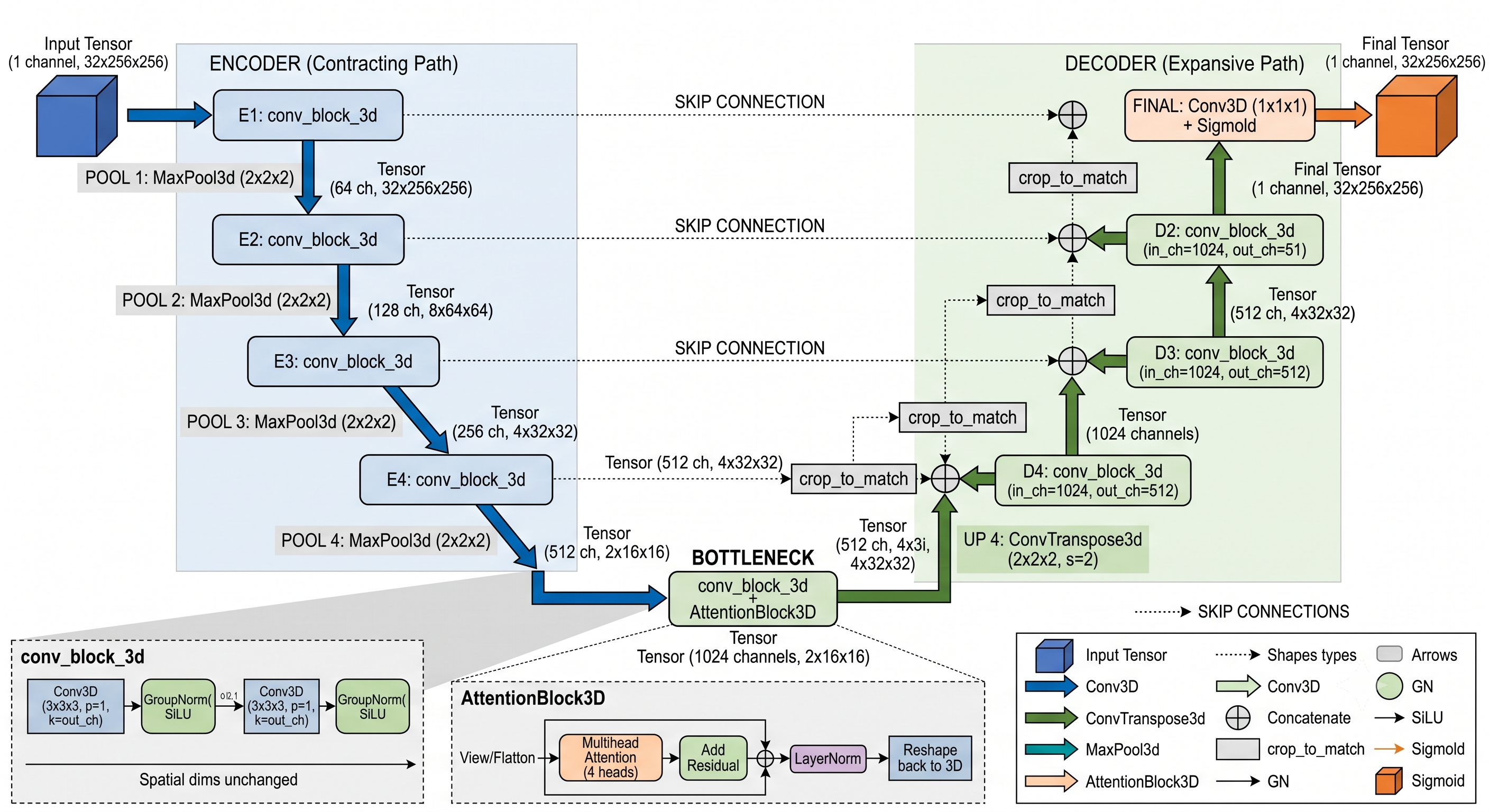}
    \caption{3D U-Net architecture}
    \label{fig:unet_arch}
\end{figure}

The regression network is based on a three-dimensional U-Net architecture (Figure~\ref{fig:unet_arch}), consisting of an encoder-decoder structure with symmetric skip connections that preserve high-resolution anatomical information during feature reconstruction \cite{ronneberger2015unet}. The encoder progressively extracts hierarchical image features through successive convolutional and down-sampling operations, while the decoder restores spatial resolution by combining high-level semantic information with fine anatomical details propagated through the skip connections. This architecture enables the network to exploit both local image texture and global anatomical context for volumetric CT-to-PET synthesis.

To further improve contextual feature representation, a multi-head self-attention module is incorporated at the bottleneck of the network \cite{vaswani2017attention}. Conventional convolutional layers primarily capture local spatial relationships, whereas self-attention models long-range dependencies across the imaging volume. This is particularly beneficial in head and neck imaging, where metabolically related structures may extend across multiple anatomical regions and cannot always be adequately represented using local receptive fields alone.

A final $1\times1\times1$ convolution follwed by a sigmoid activation produces a normalized PET volume with voxel intensity constrained to the range $[0,1]$. The resulting prediction represents the regression estimate that is subsequently combined with the adversarial pathway during hotspot-guided fusion.

\textbf{Loss Function.}
A major challenge in CT-to-PET synthesis is the severe imbalance between metabolically active tumor voxels and normal background tissue. High-SUV tumor regions typically occupy only a small fraction of the imaging volume, causing conventional voxel-wise loss functions to be dominated by background voxels. Consequently, models trained using a standard L1 or L2 loss frequently underestimate SUV within FDG-avid lesions.
To mitigate this imbalance, the regression pathway is optimized using an SUV-weighted objective function that combines voxel-wise L1 loss with structural similarity (SSIM) regularization\cite{wang2004image},

\begin{equation}
    \mathcal{L}_{\text{reg}} = \frac{1}{N}\sum_i w(y_i)\,|y_i - \hat{y}_i|
    + \lambda_{\text{SSIM}}\cdot\bigl(1 - \text{SSIM}(y,\hat{y})\bigr)
    \label{eq:loss_reg}
\end{equation}
where the spatially varying weight $w(y)$ amplifies the gradient contribution of high-SUV voxels quadratically:
\begin{equation}
    w(y) = 1 + \alpha y^{2},\qquad \alpha = 5,\quad \lambda_{\text{SSIM}} = 0.5
    \label{eq:weight}
\end{equation}
assigns progressively larger penalties to errors in voxels with higher SUV values. This weighting encourages the network to prioritize accurate prediction within metabolically active lesions while maintaining stable reconstruction of surrounding normal tissue. The SSIM term complements the voxel-wise regression loss by preserving local structural consistency and reducing excessive smoothing that commonly occurs when optimizing only pixel-wise intensity differences.

The weighted regression objective and three-dimensional U-Net architecture provide the quantitative foundation of the proposed framework, producing stable SUV estimates that are subsequently integrated with the adversarial pathway to generate the final synthetic PET volume.

\subsection{Adversarial Path}

Although the regression pathway provides stable voxel-wise standardized uptake value (SUV) estimation, optimization using pixel-wise loss functions inherently suppresses high-frequency image information, resulting in overly smooth PET reconstructions. To recover realistic PET appearance while preserving anatomical consistency with the input CT, the proposed framework incorporates a three-dimensional conditional generative adversarial network (cGAN) \cite{goodfellow2014generative, isola2017image} as a complementary synthesis pathway.

\begin{figure}[H]
    \centering
    \includegraphics[width=\textwidth]{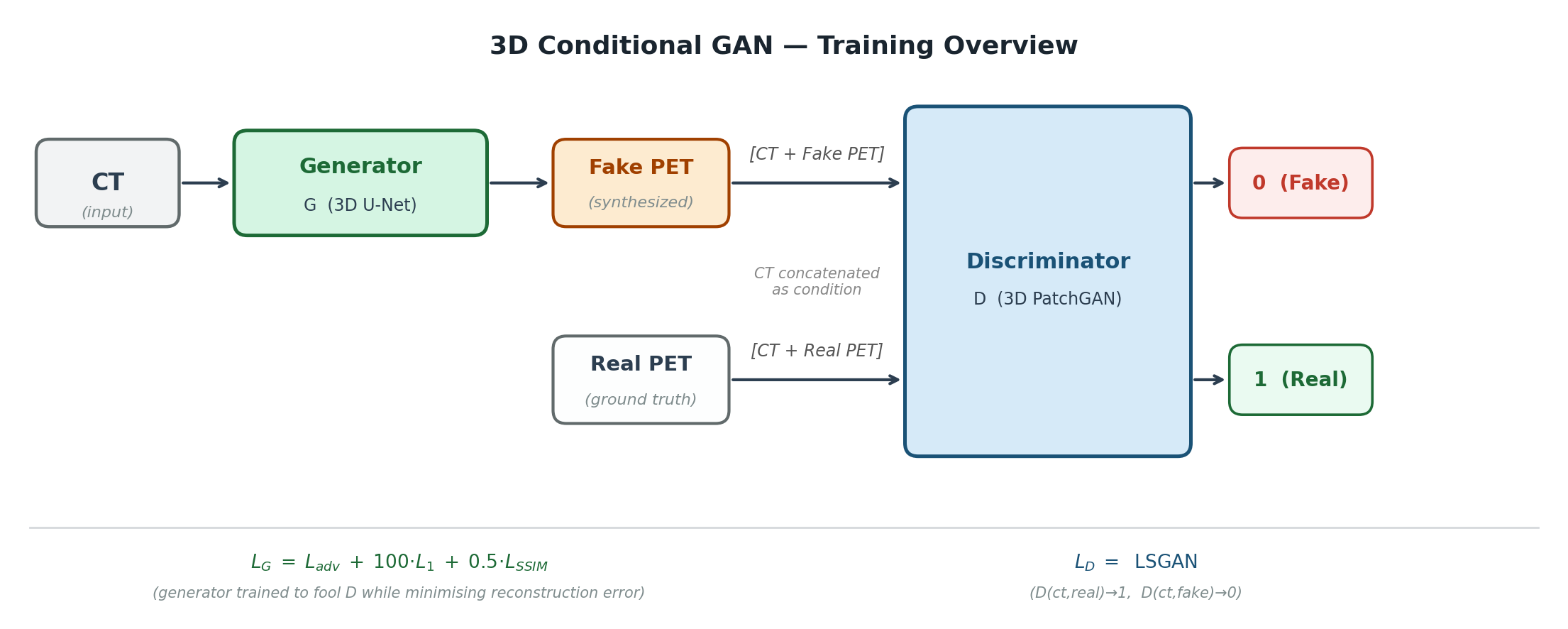}
    \caption{3D conditional Generative Adversarial Network (cGAN) training overview}
    \label{fig:cgan_overview}
\end{figure}

The adversarial pathway consists of a CT-conditioned generator and a discriminator that are trained simultaneously through adversarial learning. Rather than minimizing voxel-wise reconstruction error alone, the generator learns to synthesize PET volumes that resemble the distribution of real PET images, while the discriminator learns to distinguish synthetic PET from the corresponding reference PET. This adversarial formulation encourages realistic reconstruction of local image texture, intensity variation, and metabolic heterogeneity that are often attenuated by regression-based optimization. An overview of the adversarial training framework is shown in Figure~\ref{fig:cgan_overview}.

The generator adopts the same three-dimensional U-Net architecture described in Section~2.4. Using a common backbone ensures that differences between the regression and adversarial pathways arise from their optimization objectives rather than architectural complexity. The discriminator employs a three-dimensional PatchGAN architecture that evaluates local volumetric patches instead of assigning a single realism score to the entire image. This patch-based formulation encourages realistic reconstruction of both local PET texture and global metabolic appearance while remaining computationally efficient for volumetric image synthesis.

Although adversarial learning substantially improves perceptual realism, the discriminator evaluates image realism rather than quantitative SUV accuracy. Consequently, visually plausible PET images may still exhibit inaccurate SUV estimates within metabolically active lesions. This limitation motivates the hotspot-guided Laplacian pyramid blending strategy presented in the following section, where the complementary strengths of the regression and adversarial predictions are selectively combined to generate the final synthetic PET volume.

\textbf{Adversarial Loss Function.} 
The adversarial pathway is optimized using the Least Squares Generative Adversarial Network (LSGAN) objective \cite{mao2017least}, which replaces the binary cross-entropy formulation of conventional GANs with a least-squares loss. Compared with the original GAN objective, LSGAN provides smoother gradients, improves optimization stability, and reduces the likelihood of vanishing gradients during training. To further stabilize optimization, one-sided label smoothing is applied by assigning target values of 0.9 for real PET volumes and 0 for synthetic PET volumes.

The discriminator minimizes
\begin{equation}
    \mathcal{L}_D =
    \tfrac{1}{2}\,
    \mathbb{E}\!\left[(D(x,y)-0.9)^2\right]
    +
    \tfrac{1}{2}\,
    \mathbb{E}\!\left[D(x,\hat{y})^2\right],
    \label{eq:loss_D}
\end{equation}

where $x$ denotes the input CT volume, $y$ the reference PET volume, and $\hat{y}$ the synthetic PET prediction generated by the network.

The generator minimizes

\begin{equation}
    \mathcal{L}_G =
    \mathbb{E}\!\left[(D(x,\hat{y})-1)^2\right]
    +
    \lambda_{L1}\,
    \mathbb{E}\!\left[|y-\hat{y}|\right]
    +
    \lambda_{\mathrm{SSIM}}
    \left(
    1-\mathrm{SSIM}(y,\hat{y})
    \right),
    \label{eq:loss_G}
\end{equation}
where the adversarial loss encourages realistic PET appearance, the L1 term preserves voxel-wise quantitative fidelity, and the SSIM term maintains structural consistency with the reference PET. The weighting coefficients were empirically set to $\lambda_{L1}=100$ and $\lambda_{\mathrm{SSIM}}=0.5$, balancing quantitative accuracy and perceptual realism during optimization.

Although adversarial optimization substantially improves image realism, the discriminator evaluates perceptual appearance rather than quantitative tracer uptake. Consequently, visually realistic PET images may still underestimate SUV within metabolically active lesions. This limitation motivates the hotspot-guided Laplacian pyramid blending strategy described in the following section, where the regression and adversarial predictions are selectively blended to preserve quantitative fidelity within predicted hotspots while maintaining realistic PET texture elsewhere.

\subsection{Hotspot-Guided Laplacian Pyramid Blending}

The final synthetic PET volume is obtained by blending the regression and adversarial predictions using a hotspot-guided Laplacian pyramid blending framework. Rather than combining both predictions uniformly across the entire image, the proposed approach performs spatially adaptive blending that preferentially preserves quantitative SUV information within predicted metabolically active regions while maintaining realistic PET appearance elsewhere. The blending procedure consists of two sequential stages: (1) hotspot mask generation and (2) multiscale Laplacian pyramid blending.

\textbf{Hotspot Mask Generation.}
A spatially varying soft mask, $M\in[0,1]$, is generated from the regression prediction to identify regions where quantitative SUV information should be preferentially preserved. Since the regression pathway provides the most reliable voxel-wise SUV estimates, it serves as the basis for estimating metabolically active regions.

The hotspot mask is constructed by first applying a binary threshold of $\tau=0.125$, corresponding to an SUV of approximately 2.5, to the normalized regression prediction. The binary mask is subsequently dilated using a three-dimensional $7\times7\times7$ max-pooling kernel to account for uncertainty at lesion boundaries. Finally, a three-dimensional Gaussian smoothing kernel ($\sigma=2$) is applied to generate a continuous weighting function.

The resulting mask assigns values approaching one to predicted metabolically active regions and values approaching zero to surrounding background tissue. The smooth transition between these extremes avoids abrupt changes during blending and enables seamless integration of the regression and adversarial predictions.

\textbf{Laplacian Pyramid Blending.}
Given the regression prediction $V_{\mathrm{reg}}$, adversarial prediction $V_{\mathrm{GAN}}$, and hotspot mask $M$, the final synthetic PET volume is generated using a three-dimensional extension of the Laplacian pyramid blending framework proposed by Burt and Adelson \cite{burt1983laplacian}.

Gaussian pyramids are first constructed for both predictions and the hotspot mask through repeated Gaussian filtering and two-fold downsampling. A Gaussian pyramid of $L=4$ levels is built by repeated 3D Gaussian blur and $2\times$ subsampling. The Laplacian pyramid band-pass residuals are:
\begin{equation}
    \mathcal{L}^{(l)} = G^{(l)} - \mathrm{up}\!\left(G^{(l+1)}\right),\qquad l=0,\ldots,L-2
    \label{eq:laplacian}
\end{equation}
where $G^{(l)}$ denotes the Gaussian representation at pyramid level $l$.
At each pyramid level, the regression and adversarial coefficients are blended according to: 
\begin{equation}
    \mathcal{L}^{(l)}_{\text{blend}} = M^{(l)}\cdot\mathcal{L}^{(l)}_{\text{reg}}
    + \bigl(1-M^{(l)}\bigr)\cdot\mathcal{L}^{(l)}_{\text{GAN}}
    \label{eq:blend}
\end{equation}
where $M^{(l)}$ is the Gaussian pyramid representation of the hotspot mask. The blended Laplacian coefficients are subsequently reconstructed to obtain the final synthetic PET volume.

Unlike direct voxel-wise averaging, Laplacian pyramid blending combines information across multiple spatial scales, allowing low-frequency quantitative information from the regression prediction to be preserved while maintaining the high-frequency texture synthesized by the adversarial pathway. Because blending is guided by a continuous hotspot mask rather than a binary segmentation, transitions between the two prediction streams remain smooth, reducing boundary artifacts and producing anatomically consistent synthetic PET volumes.

\subsection{Implementation Details}
\textbf{Training Strategy.}
The proposed framework was implemented in PyTorch 2.0 and trained on the Purdue Gilbreth High Performance Computing (HPC) cluster using NVIDIA A100 GPUs (40--80 GB VRAM) with CUDA 13.1. The regression network was trained first to convergence and subsequently used to initialize the generator of the adversarial pathway. This warm-start initialization provides a quantitatively meaningful starting point for adversarial optimization, reducing unstable early training and accelerating convergence.

The regression and adversarial networks were optimized independently using the Adam optimizer with a learning rate of $2\times10^{-4}$, $\beta_1=0.5$, and $\beta_2=0.999$. The adversarial generator was optimized using the composite objective described in Equation~(\ref{eq:loss_G}), while the discriminator was optimized using Equation~(\ref{eq:loss_D}). The weighting coefficients were fixed at $\lambda_{L1}=100$ and $\lambda_{\mathrm{SSIM}}=0.5$ throughout training. Both networks were trained for 50 epochs, and the model with the lowest validation loss was retained for subsequent evaluation.

\textbf{Sliding-Window Volume Reconstruction.}
Full patient volumes were reconstructed using a sliding-window inference strategy applied to fixed-size three-dimensional patches. Adjacent volumetric patches were extracted with a 50\% overlap, corresponding to a stride of 16 axial slices. Predictions from overlapping regions were combined using cosine-weighted averaging to reduce boundary artifacts while preserving continuity across adjacent patches.

The blended patches were subsequently assembled into a complete three-dimensional synthetic PET volume, ensuring seamless transitions between neighboring patches without introducing visible stitching artifacts.
\subsection{Performance Evaluation}

The proposed framework was evaluated using complementary quantitative, image quality, and qualitative clinical assessments. Quantitative metrics were used to evaluate voxel-wise reconstruction accuracy, image quality metrics assessed structural and perceptual similarity, and qualitative clinical evaluation examined preservation of metabolically active lesions and overall anatomical realism. Performance was computed independently for each fold of the five-fold cross-validation and is reported as the mean $\pm$ standard deviation across all validation patients.

\textbf{Quantitative Accuracy.}
Voxel-wise reconstruction accuracy was evaluated using the mean absolute error (MAE) between the synthesized PET volume and the corresponding reference PET volume,

\begin{equation}
\mathrm{MAE}
=
\frac{1}{N}
\sum_{i=1}^{N}
|y_i-\hat{y}_i|,
\end{equation}

where $y_i$ and $\hat{y}_i$ denote the reference and synthesized PET voxel intensities, respectively, and $N$ is the total number of voxels.

Because accurate quantification of tracer uptake is essential for clinical PET interpretation, standardized uptake value (SUV) measurements were also evaluated. Quantitative agreement between the synthesized and reference PET images was assessed using SUV-based metrics within metabolically active regions, with particular emphasis on preserving high-uptake lesions identified by the reference PET images.

\textbf{Image Quality.}
Image quality was assessed using the structural similarity index measure (SSIM) and the peak signal-to-noise ratio (PSNR), which evaluate complementary aspects of image fidelity.

SSIM quantifies structural similarity by jointly evaluating luminance, contrast, and structural information between the synthesized and reference PET images. Higher SSIM values indicate greater structural agreement and improved preservation of anatomical detail.

PSNR measures the reconstruction quality relative to the voxel-wise error between synthesized and reference PET images. Higher PSNR values indicate reduced reconstruction error and improved overall image fidelity.

These metrics (MAE, SSIM, and PSNR) provide complementary assessments of quantitative accuracy, structural preservation, and perceptual image quality.

\textbf{Clinical Evaluation.}
Quantitative image similarity does not necessarily guarantee preservation of clinically relevant metabolic information. Therefore, qualitative clinical evaluation was performed by visually comparing the synthesized PET images with the corresponding reference PET images.

Representative cases were examined to assess lesion conspicuity, preservation of metabolically active hotspots, and anatomical consistency throughout the reconstructed volumes. Particular attention was given to the localization of regions exhibiting elevated FDG uptake and the ability of the proposed hotspot-guided Laplacian pyramid blending strategy to preserve quantitative lesion characteristics while maintaining realistic PET appearance.

Qualitative comparisons between the regression, adversarial, blended, and reference PET images were included to illustrate the complementary contributions of each synthesis pathway and the effectiveness of the proposed blending strategy.

\section{Results}
\label{sec:results}

The proposed dual-path framework was evaluated using five-fold patient-level cross-validation on the 44-patient QIN-HEADNECK dataset. Quantitative performance was assessed on reconstructed three-dimensional PET volumes using the evaluation metrics described in Section~2.8, while qualitative analysis examined lesion localization and overall image appearance. Training convergence, quantitative performance, and representative clinical examples are presented in the following sections.

Table~\ref{tab:perfold} summarizes the quantitative performance obtained for each cross-validation fold. Both the regression and adversarial pathways demonstrated consistent performance across all five folds, with only modest variation in MAE, PSNR, and SSIM-W. The regression pathway achieved a mean MAE of $0.0048 \pm 0.0006$, PSNR of $37.6 \pm 0.8$ dB, and SSIM-W of $0.959 \pm 0.010$, while the adversarial pathway achieved a mean MAE of $0.0045 \pm 0.0005$, PSNR of $37.5 \pm 0.6$ dB, and SSIM-W of $0.961 \pm 0.010$. The small standard deviations across folds indicate that model performance was reproducible and not dependent on a particular patient split, supporting the robustness of the proposed training strategy.

\begin{figure}[t]
  \centering
  \includegraphics[width=0.85\linewidth]{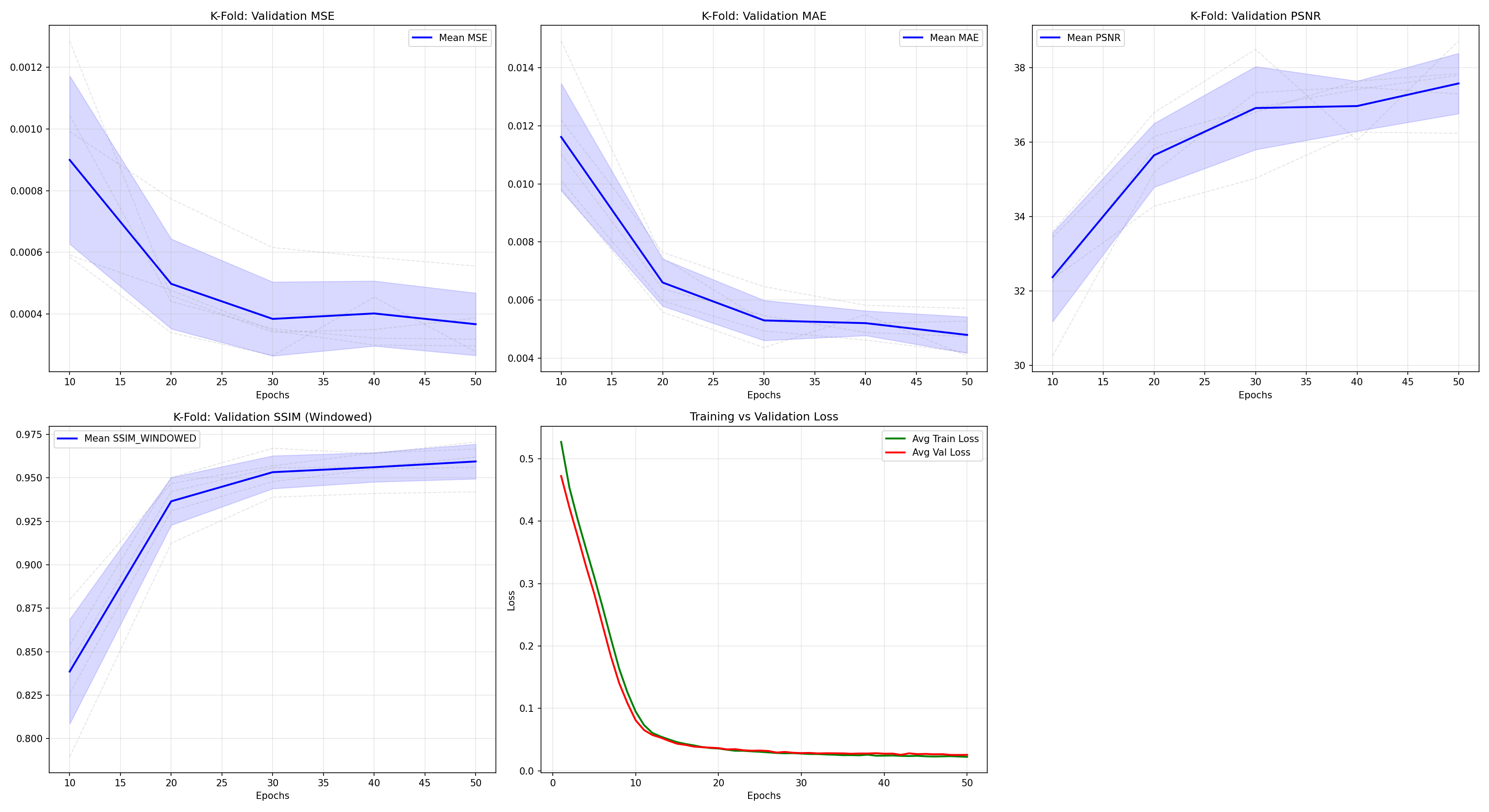}
  \caption{U-Net validation loss curves across five folds (50 epochs each).}
  \label{fig:unet_loss}
\end{figure}

\subsection{Training Convergence and Cross-Validation Stability}
\label{sec:convergence}

\begin{figure}[t]
  \centering
  \includegraphics[width=0.85\linewidth]{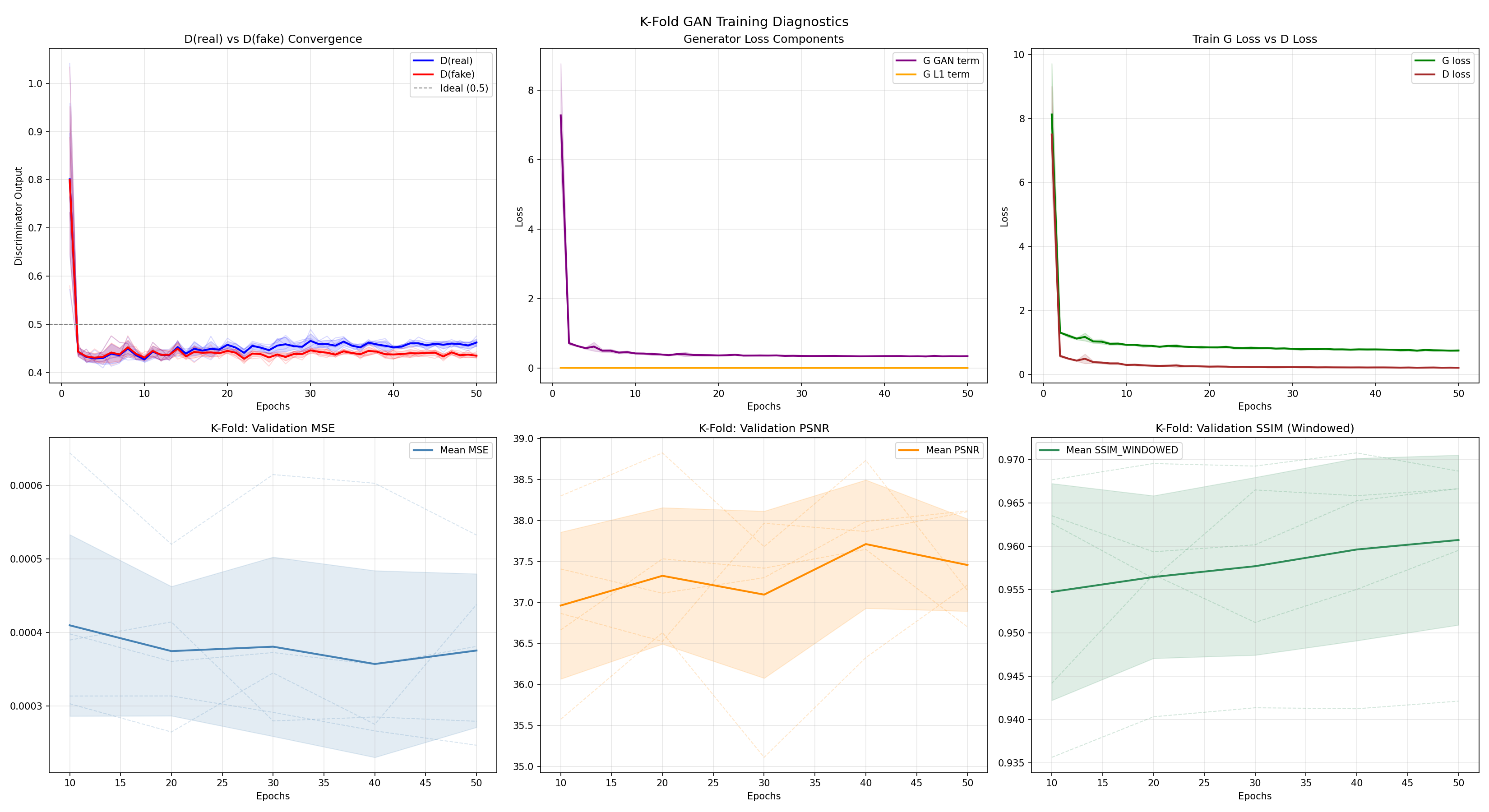}
  \caption{GAN validation loss curves across five folds. Discriminator loss stabilizes near 0.5; Fold 4 achieves the lowest validation loss (0.0165).}
  \label{fig:gan_loss}
\end{figure}

Both networks were trained for 50 epochs per fold under 5-fold cross-validation, with roughly 35 to 36 patients used for training and 8 to 9 held out for validation in each fold. Figures~\ref{fig:unet_loss} and~\ref{fig:gan_loss} show the validation loss curves for all five folds.

Table~\ref{tab:perfold} lists the per-fold validation metrics at epoch 50. Both models settle down by epoch 30 and stay stable through epoch 50. The GAN discriminator loss holds near 0.5 across every fold, which is what a healthy adversarial equilibrium looks like rather than mode collapse or a discriminator running away with training. The standard deviations are tight, with PSNR SD of $\pm 0.81$~dB for the U-Net and $\pm 0.57$~dB for the GAN, so performance doesn't swing much from one patient split to the next, and no single easy or hard fold is carrying the average. The GAN also beats the U-Net on MAE in four of the five folds, so its quantitative edge is consistent across the cross-validation and isn't tied to any one split.

\begin{table}[t]
\centering
\caption{Per-fold validation metrics at epoch 50. MAE on normalized SUV $\in [0,1]$.}
\label{tab:perfold}
\resizebox{\linewidth}{!}{%
\begin{tabular}{c ccc ccc}
\hline
& \multicolumn{3}{c}{Regression U-Net} & \multicolumn{3}{c}{Conditional GAN} \\
Fold & MAE $\downarrow$ & PSNR $\uparrow$ & SSIM-W $\uparrow$ & MAE $\downarrow$ & PSNR $\uparrow$ & SSIM-W $\uparrow$ \\
\hline
1 & 0.005274 & 37.29 & 0.9420 & 0.004888 & 36.70 & 0.9421 \\
2 & 0.005712 & 36.24 & 0.9560 & 0.005105 & 37.21 & 0.9595 \\
3 & 0.004193 & 37.79 & 0.9619 & 0.003529 & 38.12 & 0.9666 \\
4 & 0.004079 & 38.71 & 0.9706 & 0.004646 & 37.15 & 0.9687 \\
5 & 0.004746 & 37.83 & 0.9665 & 0.004496 & 38.11 & 0.9666 \\
\hline
Mean $\pm$ SD & $0.0048 \pm 0.0006$ & $37.6 \pm 0.8$ & $0.959 \pm 0.010$ & $0.0045 \pm 0.0005$ & $37.5 \pm 0.6$ & $0.961 \pm 0.010$ \\
\hline
\end{tabular}%
}
\end{table}

\subsection{Quantitative Accuracy}
\label{sec:quant_accuracy}

Table~\ref{tab:fullcohort} reports the full-cohort performance of all three outputs, the regression U-Net, the conditional GAN, and the blend, evaluated on all 44 patients as per-patient mean $\pm$ standard deviation. We break these results down by metric below.

\begin{table}[t]
\centering
\caption{Full-cohort performance. The Fold-4 model for each pathway was evaluated on all 44 patients; values are the per-patient mean $\pm$ standard deviation.}
\label{tab:fullcohort}
\begin{tabular}{lccc}
\hline
Method & MAE $\downarrow$ & PSNR (dB) $\uparrow$ & SSIM-W $\uparrow$ \\
\hline
Regression U-Net & $0.004329 \pm 0.001622$ & $38.71 \pm 2.20$ & $0.9613 \pm 0.0399$ \\
Conditional GAN  & $0.003841 \pm 0.001539$ & $39.29 \pm 2.03$ & $0.9636 \pm 0.0399$ \\
Blend (3D)       & $0.003946 \pm 0.001659$ & $39.19 \pm 2.24$ & $0.9634 \pm 0.0407$ \\
\hline
\end{tabular}
\end{table}

\subsubsection{Mean Absolute Error (MAE)}
\label{sec:mae}

MAE a direct measure of voxel-wise error, so a lower value means the synthetic volume tracks the real PET intensities more closely on average. Averaged over all 44 patients, the conditional GAN reaches an MAE of $0.003841 \pm 0.001539$ and the blend $0.003946 \pm 0.001659$, with the regression U-Net a little behind at $0.004329 \pm 0.001622$. On paper the GAN edges out the blend, but that gap of about $0.0001$ is tiny next to the patient-to-patient spread of roughly $0.0015$. In other words the GAN and the blend are statistically tied on global MAE, and that's exactly the result we want. The blend isn't trying to beat the GAN on the whole-image average. It keeps the GAN's global quality everywhere except inside the predicted high-uptake mask, which covers only about 1.5 to 5\% of the voxels, and inside that mask it swaps in the more reliable regression SUV. So the blend buys better-behaved tumor uptake at essentially no cost to global accuracy, which is the whole reason we fuse the two paths instead of just using the GAN. On the full-volume metrics the conditional GAN edges out the regression U-Net (PSNR $\Delta = +0.58$~dB, MAE $\Delta = -11.3\%$), consistent with the discriminator penalizing blur and enforcing realistic texture.

\subsubsection{SUV Accuracy}
\label{sec:suv_accuracy}
SUV, the standardized uptake value, is the clinical unit radiologists actually read PET images in, and it's normalized here to $[0,1]$ for network training. Unlike the global metrics above, SUV accuracy in this study isn't a separate computed statistic. It's assessed qualitatively within the metabolically active tumor region, since that's where getting the tracer uptake right matters most for clinical reading. So we paid closest attention to SUV agreement inside the high-uptake lesions the reference PET flags, rather than to the volume-wide average. The regression U-Net is the weakest output on the global metrics, but it's the most reliable one in the high-uptake tumor regions, and that's exactly why it drives the blending mask: the mask itself is built from a threshold on the regression output (SUV $\approx 2.5$), so wherever the regression path identifies elevated uptake, the blend preserves its SUV rather than the GAN's. All three outputs still underestimate SUV in the most intensely FDG-avid foci, which lines up with what earlier CT-to-PET studies have reported. Because the blend keeps the regression's SUV inside those hotspots, it eases that underestimation compared to the GAN alone, though it doesn't erase it.

\begin{figure}[H]
    \centering
    \includegraphics[width=\textwidth]{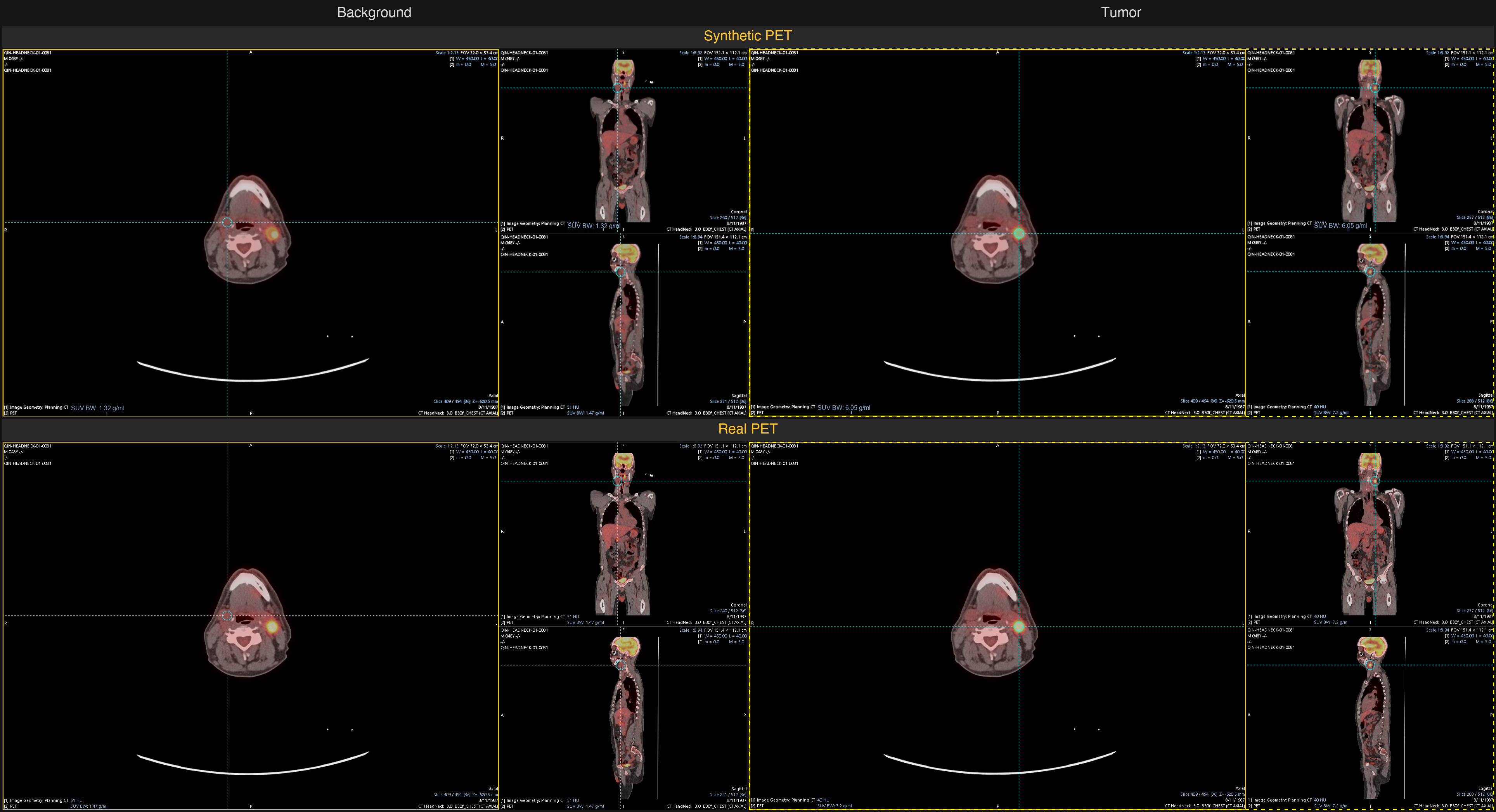}
    \caption{Patient 0081: synthetic PET (top) vs.\ real PET (bottom). Left: background; right: tumor region. Lesion is correctly localized but synthetic SUV is visibly underestimated.}
    \label{fig:suv_underest}
\end{figure}

Across the cohort, the synthetic outputs underestimated SUV within the gross tumor volume by roughly 15--20\%, even though the FDG-avid lesions were generally localized correctly. Figure~\ref{fig:suv_underest} illustrates this for a representative patient: the lesion is localized accurately, but the synthetic SUV in the tumor core is visibly lower than the ground-truth PET.

\subsection{Image Quality}
\label{sec:image_quality}

\subsubsection{PSNR}
\label{sec:psnr}
PSNR is more sensitive to large, localized errors than MAE is, so a higher PSNR means fewer big voxel-wise mistakes anywhere in the volume. The blend reaches $39.19 \pm 2.24$~dB and the GAN $39.29 \pm 2.03$~dB, with the regression U-Net at $38.71 \pm 2.20$~dB. All three sit within a fraction of a dB of each other, and well inside the roughly $\pm 2$~dB spread across patients, so on reconstruction fidelity they're effectively even. The tight clustering tells us none of the three is trading away image quality.

\subsubsection{SSIM}
\label{sec:ssim}
SSIM compares luminance, contrast, and structure between two images rather than just raw intensity differences, and we compute it in a windowed fashion (SSIM-W) over local patches across the volume. A higher SSIM-W means the synthetic PET preserves the same local structure and contrast pattern as the real scan, not just similar average brightness. SSIM-W is high and nearly identical across the board: $0.9634 \pm 0.0407$ for the blend, $0.9636 \pm 0.0399$ for the GAN, and $0.9613 \pm 0.0399$ for the regression U-Net. Numbers this high, and this close, mean the outputs all hold onto local structure and contrast relative to the real PET, and that folding in the regression SUV inside the hotspots doesn't disturb the structural agreement the GAN already provides.

\subsection{Clinical Evaluation}
\label{sec:clinical_eval}

\subsubsection{Qualitative Comparison}
\label{sec:qualitative}

\begin{figure}[t]
  \centering
  \includegraphics[width=0.85\linewidth]{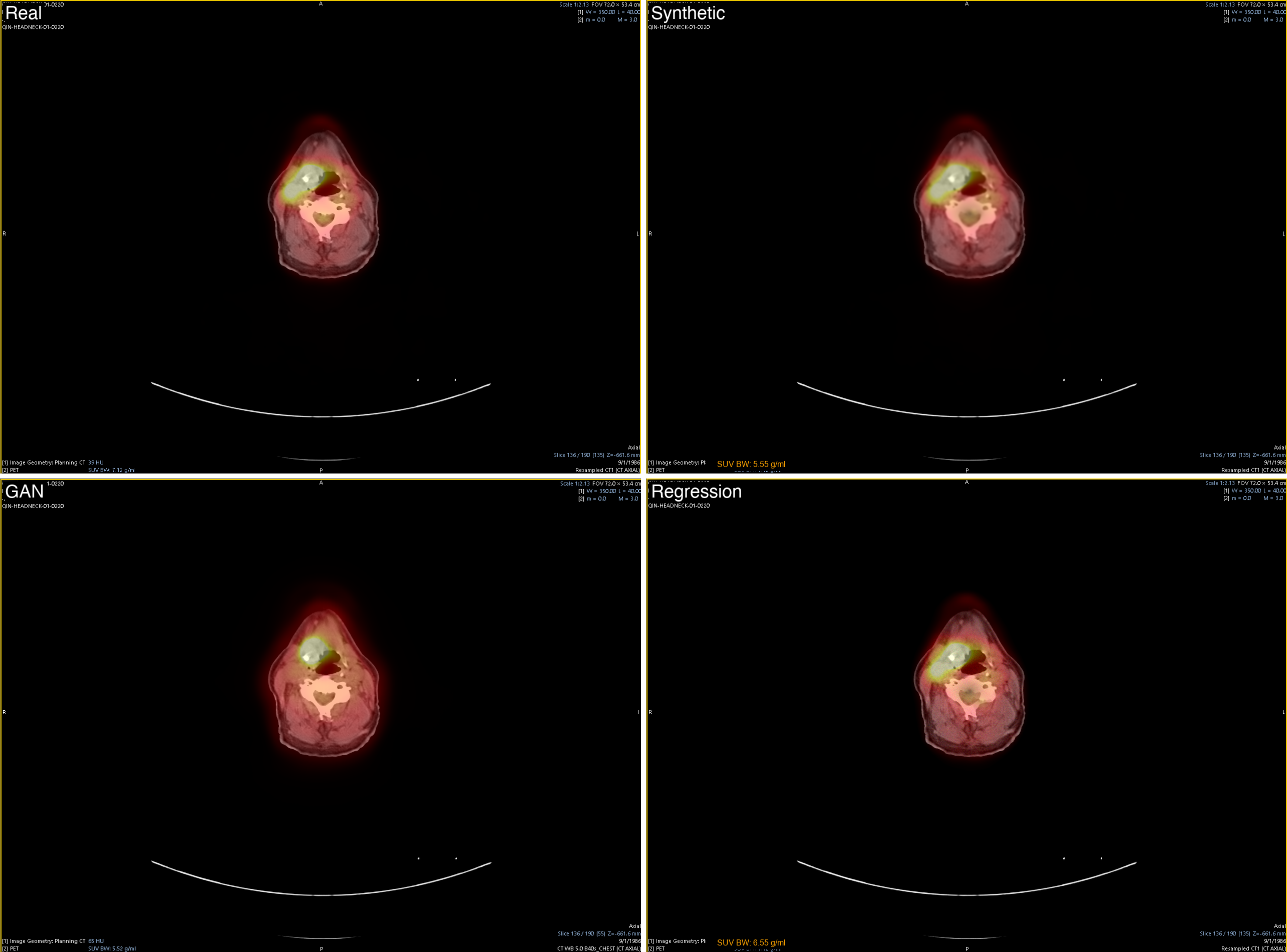}
  \caption{Patient 0220: regression, GAN, and blended synthetic PET vs. real
PET}
  \label{fig:p0220}
\end{figure}

Representative examples of the regression, adversarial, blended, and reference PET volumes are shown in Figures~\ref{fig:p0220}--\ref{fig:p0195}. Across all evaluated cases, the regression pathway produced smoother images with relatively homogeneous uptake patterns, whereas the adversarial pathway generated sharper PET texture and improved visual realism. The blended output combined these complementary characteristics by preserving realistic background appearance while maintaining lesion conspicuity and overall metabolic distribution.

Patient 0220 (Figure~\ref{fig:p0220}) shows all three outputs against the real PET. The regression output puts the hotspot in the right anatomical spot with reasonable SUV values, but its edges are soft and the surrounding tissue looks blurry and uniform. The GAN output looks a lot more like real PET, with convincing background texture, but the hotspot's extent and intensity drift away from the ground truth and faint uptake shows up in places that shouldn't be lighting up. The blend gets the best of both: controlled SUV in the hotspot like the regression, GAN-like texture in the background, and a clean transition across the mask boundary. Of the three, it's the closest visual match to the real PET.

\subsubsection{Lesion Localization}
\label{sec:localization}

\begin{figure}[t]
  \centering
  \includegraphics[width=0.85\linewidth]{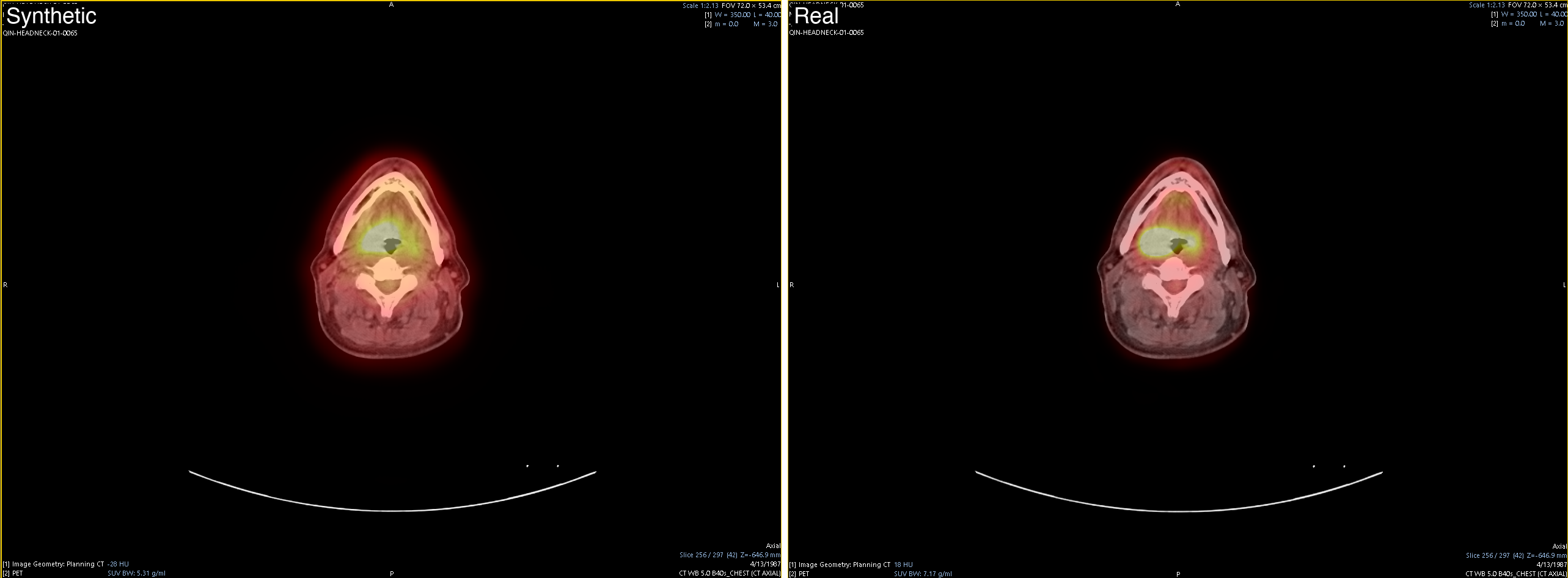}
  \caption{Patient 0065: blended synthetic PET vs. real PET}
  \label{fig:p0065}
\end{figure}

Patient 0065 (Figure~\ref{fig:p0065}) shows the blended synthetic PET against the real PET. The FDG-avid tumor is localized correctly and the hotspot's extent roughly tracks the real PET. You can see the SUV underestimation, since the synthetic hotspot isn't as intensely colored as the ground truth, but the localization is good enough for the triage job we have in mind: it flags that this patient has a suspicious high-uptake region worth a real PET scan.

\begin{figure}[t]
  \centering
  \includegraphics[width=0.85\linewidth]{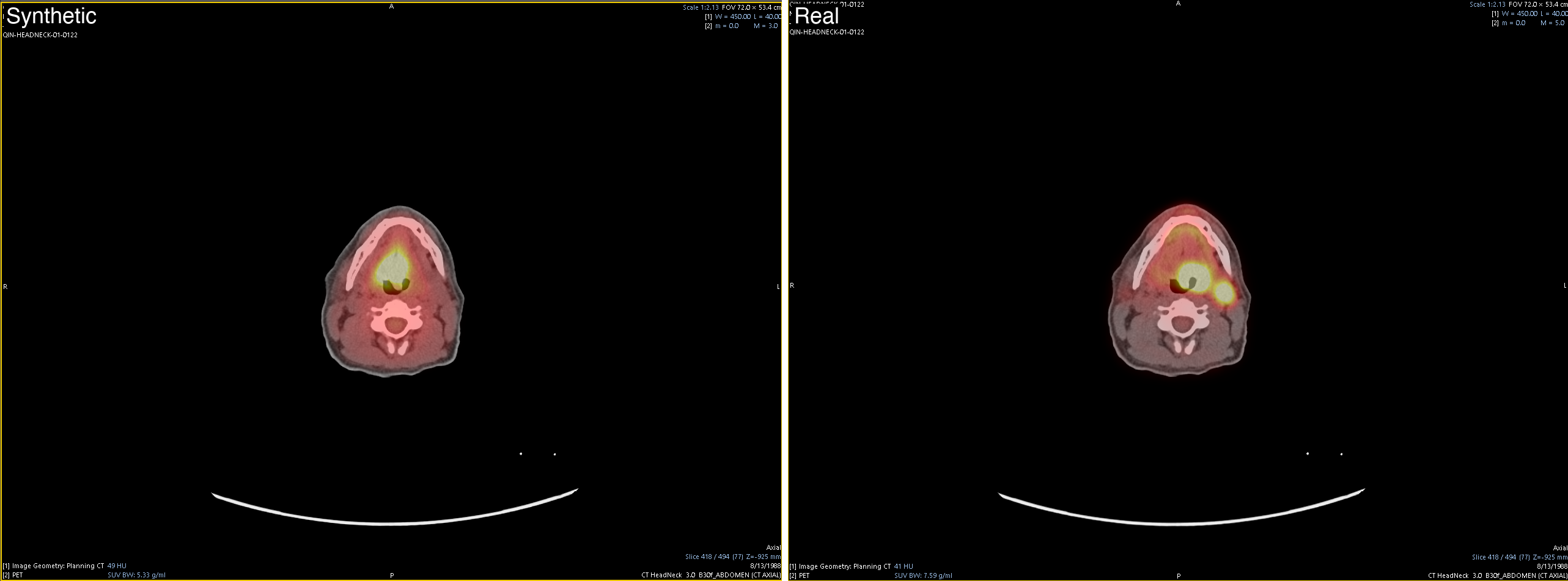}
  \caption{Patient 0122: blended synthetic PET vs. real PET}
  \label{fig:p0122}
\end{figure}

Patient 0122 (Figure~\ref{fig:p0122}) is the harder case. The hotspot is close but not exact, and it's shifted relative to the ground truth. This is where CT-alone synthesis runs into its limits in head and neck, because the CT signal doesn't always mark the tumor boundary clearly enough for the model to pin it down with confidence. It's exactly why we frame synthetic PET as a triage tool and not a diagnostic replacement. The predicted region is useful as a flag to go investigate with a real scan, but it shouldn't be used to read SUV\textsubscript{max} or make clinical calls.

\subsubsection{Hotspot Preservation}
\label{sec:hotspot_preservation}

\begin{figure}[t]
  \centering
  \includegraphics[width=0.85\linewidth]{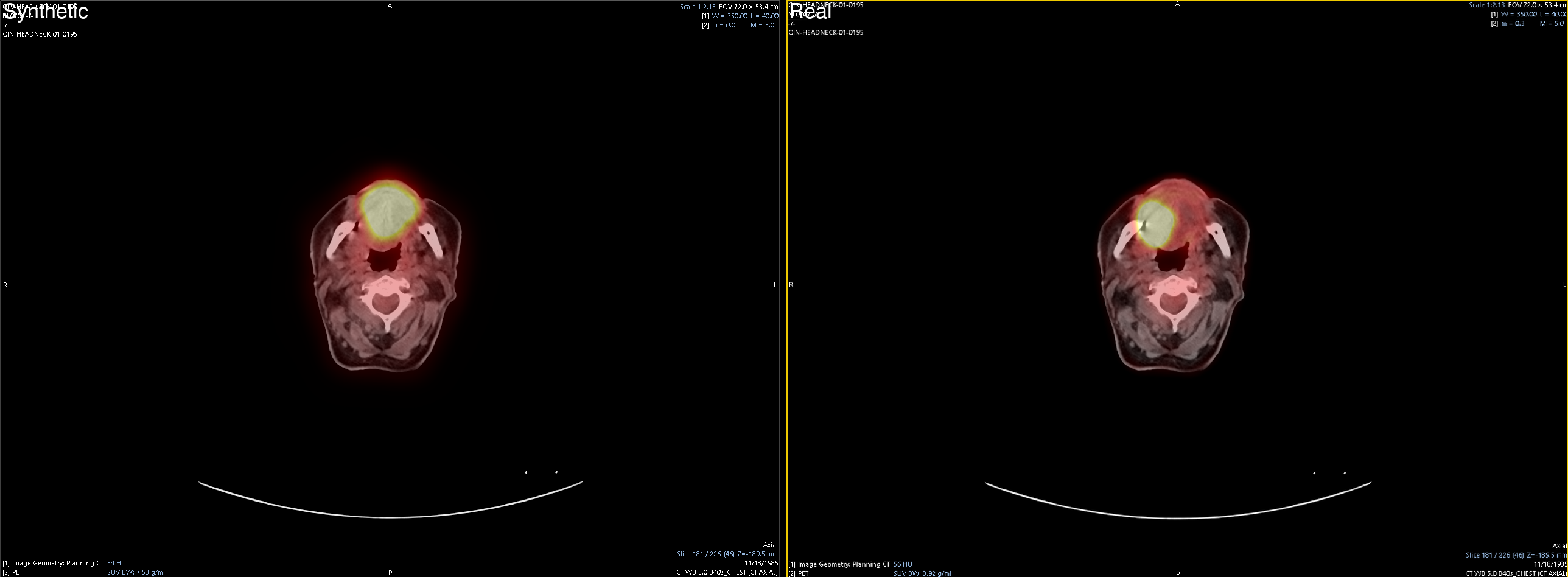}
  \caption{Patient 0195: blended synthetic PET vs. real PET}
  \label{fig:p0195}
\end{figure}

Patient 0195 (Figure~\ref{fig:p0195}) is a partial-localization miss. The synthetic PET does convey a high-avidity region in the right general area, but it doesn't fully recover the tumor's extent or boundary, and the predicted hotspot comes out smaller and less defined than the ground truth. The signal is still strong enough to say metabolic activity is present, which fits the triage framing, but it's a reminder that spatial accuracy drops off when CT contrast is weakest. Across the cohort, wherever the regression pathway localized a lesion, the blend held onto that high-uptake region rather than washing it out toward the smoother GAN estimate, which is the whole point of the Laplacian pyramid fusion: it keeps the quantitative hotspot content instead of averaging it away.

\section{Discussion}
\textbf{Principal Findings}
The present study demonstrates that quantitative SUV fidelity and perceptual image realism need not be treated as competing objectives in CT-to-PET synthesis. Instead, our results show that these complementary objectives can be optimized independently and subsequently reconciled through anatomically informed hotspot-guided fusion. Using this strategy, the proposed framework generated synthetic PET images that preserved metabolically relevant quantitative information while maintaining realistic PET appearance, thereby addressing one of the longstanding challenges in synthetic PET generation.

This finding extends beyond the specific architecture presented in this work. Rather than demonstrating only the effectiveness of a dual-path network, our results suggest a broader conceptual framework for quantitative medical image synthesis. By separating quantitative reconstruction from perceptual synthesis before spatially adaptive integration, the proposed approach provides a practical solution for balancing quantitative accuracy and image realism, two objectives that have traditionally been optimized simultaneously within a single network.

\textbf{Interpretation of the Dual-Path Framework} 
Pixel-wise regression and adversarial learning represent two fundamentally different optimization strategies for medical image synthesis. Regression-based models trained with voxel-wise losses such as L1 and L2 generally achieve stable quantitative reconstruction but tend to produce overly smooth images because optimization converges toward the conditional mean of the training distribution, suppressing high-frequency image information despite minimizing reconstruction error \cite{isola2017image, johnson2016perceptual, zhao2017loss}.

Conversely, adversarial learning encourages generated images to match the statistical distribution of real PET images, substantially improving perceptual realism and local image texture. However, because the discriminator evaluates image realism rather than voxel-wise quantitative agreement, adversarial optimization alone cannot guarantee accurate SUV estimation, particularly within metabolically active lesions where quantitative PET measurements are clinically important \cite{mao2017least, wang2025pseudo,wang2018pet3dcgan, nie2020adversarial}.

Our findings are consistent with these observations but further demonstrate that the apparent conflict between quantitative SUV fidelity and perceptual realism arises primarily from attempting to optimize both objectives within a single network. By assigning each objective to an independent learning pathway, the proposed framework allows each model to specialize in its respective task before integrating both predictions through hotspot-guided Laplacian pyramid blending. Rather than performing simple voxel-wise averaging, the fusion strategy selectively preserves regression-derived quantitative information within predicted metabolically active regions while incorporating adversarially generated high-frequency texture throughout the remaining anatomy. Consequently, the improved synthetic PET images arise not simply from increased architectural complexity, but from reformulating CT-to-PET synthesis as two complementary optimization problems that are solved independently before being reconciled through anatomically informed fusion.

\textbf{Comparison with Previous CT-to-PET Synthesis Methods}
Deep learning approaches for CT-to-PET synthesis have progressed from early two-dimensional convolutional and conditional generative adversarial networks to fully volumetric architectures and, more recently, diffusion-based generative models. Table~\ref{tab:literature} summarizes representative deep learning approaches for CT-to-PET synthesis and places the proposed method within the current methodological landscape. Although direct quantitative comparison should be interpreted with caution because studies differ in datasets, preprocessing pipelines, normalization methods, and evaluation protocols, several consistent trends emerge. Early CT-to-PET synthesis studies demonstrated the feasibility of estimating metabolic information from anatomical CT using two-dimensional convolutional networks and conditional GANs, while more recent work has shifted toward fully three-dimensional architectures and diffusion-based generative models to improve volumetric consistency and image realism \cite{bencohen2019cross,chandrashekar2022virtual,nguyen2025ct}.

Despite these advances, a common challenge remains the balance between quantitative SUV fidelity and perceptual realism. Chandrashekar et al.~\cite{chandrashekar2022virtual} reported systematic SUV underestimation on the same QIN-HEADNECK dataset, while Ben-Cohen et al.~\cite{bencohen2019cross} demonstrated that regression and adversarial learning provide complementary information for PET synthesis. Our findings support these observations but further suggest that the trade-off is better addressed by separating quantitative reconstruction and perceptual synthesis into independent optimization pathways before integrating their strengths through hotspot-guided multiscale fusion. Unlike Bi et al.~\cite{bi2017synthesis}, who relied on manually delineated tumor annotations during inference, the proposed approach derives hotspot localization automatically from the regression prediction, eliminating the need for manual lesion annotation.

Recent diffusion-based methods have established new benchmarks for CT-to-PET synthesis using large multicenter datasets \cite{nguyen2025ct,mahdi2026ct}. Rather than competing directly with a particular image generator, the present study introduces a complementary optimization strategy that is independent of the underlying generative model. By demonstrating that quantitative reconstruction and perceptual synthesis can be optimized separately and subsequently reconciled through anatomically informed fusion, this work provides a conceptual framework that may be incorporated into future GAN-, diffusion-, or transformer-based PET synthesis methods. Standardized benchmark datasets and harmonized evaluation protocols will ultimately be essential for objective comparison and clinical translation of synthetic PET.

\begin{table}[htbp]
    \centering
    \small
    \caption{Comparison with CT$\to$PET synthesis literature. PSNR/SSIM shown where reported on full volumes; N/R = not reported. MAE units differ across studies.}
    \label{tab:literature}
    \resizebox{\textwidth}{!}{%
    \begin{tabular}{lcccccl}
        \hline
        \textbf{Method} & \textbf{Anatomy} & \textbf{$N$} & \textbf{Dim} & \textbf{PSNR (dB)} $\uparrow$ & \textbf{SSIM} $\uparrow$ & \textbf{Primary Metric Reported} \\
        \hline
        Chandrashekar et al.\ \cite{chandrashekar2022virtual} (cGAN)  & Head \& Neck & N/R  & 2D & N/R  & N/R    & Qualitative only; SUV underestimation noted \\
        Ben-Cohen et al.\ \cite{bencohen2019cross} (FCN + GAN)       & Liver        & 60   & 2D & N/R  & N/R    & TPR = 92.3\%, FPR = 0.25/case (lesion detection) \\
        Bi et al.\ \cite{bi2017synthesis} (multi-channel GAN)            & Lung         & 50   & 2D & N/R  & N/R    & MAE, PSNR on tumor ROI only \\
        Salehjahromi et al.\ \cite{salehjahromi2024synthetic}            & Lung         & 1478 & 3D & N/R  & N/R    & Clinical staging agreement, AUC \\
        Nguyen et al.\ \cite{nguyen2025ct} (CPDM)                 & Multi        & N/R  & 3D & 29.68 & 0.9396 & MAE = 284.61 Bq/mL (raw SUV units) \\
        3D Latent Diffusion \cite{mahdi2026ct}                & Head \& Neck & N/R  & 3D & 32.64 & 0.86   & MAE = 303.05 Bq/mL (raw SUV units) \\
        \hline
        \textbf{Ours (Blend, 3D)}                                & Head \& Neck & 44   & 3D & \textbf{39.19} & $\mathbf{0.9634}^*$ & MAE = 0.003946 (normalized SUV $\in[0,1]$) \\
        \hline
    \end{tabular}%
    }
\end{table}

\textbf{Clinical Relevance and Potential Applications} 
Although the proposed framework is not intended to replace diagnostic FDG-PET/CT, it demonstrates that clinically meaningful metabolic information can be inferred from routinely acquired planning CT. By combining the quantitative reliability of the regression pathway with the realistic image appearance of the adversarial pathway, the hotspot-guided fusion strategy consistently preserved lesion localization while improving SUV fidelity within metabolically active regions compared with either pathway alone. These findings suggest that separating quantitative reconstruction from perceptual synthesis provides a practical strategy for generating synthetic PET images that are more suitable for clinically relevant interpretation.

The proposed framework should therefore be viewed as a complementary decision-support tool rather than a substitute for diagnostic PET. Potential applications include highlighting suspicious FDG-avid regions on planning CT, supporting biologically informed target review, and facilitating future research in PET-guided radiotherapy and adaptive treatment planning. However, the observed SUV underestimation indicates that synthetic PET should not be used for quantitative clinical decision-making or treatment prescription. Instead, its greatest near-term value lies in augmenting anatomical CT with complementary functional information, particularly in settings where PET imaging is unavailable or cannot be readily repeated.

\textbf{Head and Neck vs.\ Other Anatomies} The head and neck region presents synthesis challenges absent in liver \cite{bencohen2017virtual} or lung \cite{salehjahromi2024synthetic} settings. For instance, the brain exhibits intense physiological FDG uptake immediately superior to the target area, while benign structures like the salivary glands, tonsils, and reactive lymph nodes show naturally elevated SUVs that are indistinguishable from true pathology on CT alone. Furthermore, the intricate interleaving of bone, air, and soft tissue produces highly variable CT-to-PET mappings across the patient cohort.

\textbf{Limitation and Future Directions} 
Despite the encouraging results, several limitations should be acknowledged. First, the proposed framework was developed and evaluated using a relatively small, single-institution public dataset, and its generalizability to other patient populations, scanners, and imaging protocols remains to be established through external multi-institutional validation. Second, residual SUV underestimation persisted within regions of intense FDG uptake, consistent with previous observations \cite{soret2007partial} in CT-to-PET synthesis studies, including those performed on the QIN-HEADNECK dataset \cite{chandrashekar2022virtual} . This behavior likely reflects the combined effects of class imbalance, dynamic range compression during normalization, and the inherent inability of anatomical CT to fully capture the biological processes governing FDG uptake.Specifically, high-SUV tumor voxels constitute only about 1--5\% of the volume, so even the SUV-weighted loss is pulled toward the background mean; rescaling SUV to $[0,1]$ further compresses these sparse high values into a narrow band, and the final sigmoid activation saturates before reaching the extremes, both biasing predictions away from peak uptake.

Future work should therefore focus on validating the proposed framework across larger and more diverse datasets, incorporating uncertainty estimation to improve confidence in voxel-wise predictions, and investigating lesion-aware optimization strategies and multimodal inputs to further improve quantitative SUV recovery. In addition, evaluating the proposed hotspot-guided fusion strategy within emerging diffusion- and transformer-based generative frameworks may help determine whether the conceptual advantages demonstrated in this study extend beyond the current dual-path architecture.

\section{Conclusion}

This study presented a fully three-dimensional dual-path framework for CT-to-PET synthesis in head and neck squamous cell carcinoma. The proposed approach demonstrates that quantitative SUV fidelity and perceptual image realism need not be treated as competing objectives but can instead be optimized independently and integrated through anatomically informed hotspot-guided Laplacian pyramid blending. This strategy produced synthetic PET images that more effectively preserved clinically relevant metabolic information than either the regression or adversarial pathway alone.

Although the proposed framework is not intended to replace diagnostic FDG-PET/CT, it establishes a practical framework for incorporating functional image synthesis into precision radiotherapy workflows. By demonstrating that complementary learning objectives can be reconciled through spatially adaptive fusion, this work advances CT-to-PET synthesis beyond conventional single-objective optimization and provides a foundation for future development of quantitative functional image synthesis. Continued validation in larger multi-institutional cohorts and integration with uncertainty-aware and next-generation generative models will be important steps toward translating synthetic PET into a clinically useful decision-support tool for precision radiation oncology.

\bibliographystyle{unsrt}
\bibliography{references}

\end{document}